\documentclass[twocolumn,prl,aps]{revtex4}
\usepackage{graphicx}

\begin{document}

\title{Covariant vs. non-covariant quantum collapse: Proposal for an experimental test}

\author{Antoine Suarez}
\address{Center for Quantum Philosophy \\ Berninastrasse 85, 8057 Zurich, Switzerland\\
suarez@leman.ch, www.quantumphil.org}

\date{January 16, 2014}

\begin{abstract}

Two alternative interpretations of the quantum collapse are proposed: a time-ordered and a timeless one. The time-ordered interpretation implies that the speed of light can be defined in an absolute way, while the timeless quantum collapse implies relativity and can be reckoned as covariant too. An experiment is proposed to decide between these two interpretations, which may also be considered a test of Bohm's ``preferred frame" assumption.

\end{abstract}


\maketitle

\ \\
\textbf{Introduction}.\textemdash A main postulate of standard quantum mechanics is that the decision of the outcome happens at the moment of detection (``wavefunction collapse''). This implies a \emph{nonlocal} coordination of the detectors, which cannot be explained by influences propagating with velocity $v\leq c$. So early as 1927, at the 5$^{th}$ Solvay conference, Einstein objected to this postulate by means of a \emph{single-particle} gedanken-experiment. The quantum collapse, he argued, involves ``an entirely peculiar mechanism of action at a distance, which [...] implies to my mind a contradiction with the postulate of relativity." \cite{bv}.

Astonishingly Einstein's gedanken-experiment in 1927 has been first realized in 2012 by the single-photon experiment presented in \cite{Guerreiro12}. This experiment demonstrates nonlocally coordinated detector's behavior, and also highlights something Einstein did not mention: Nonlocality is necessary to preserve such a fundamental principle as energy conservation.\cite{Guerreiro12}

\begin{figure}[t]
\includegraphics[trim = 10mm 182mm 10mm -6mm, clip, width=0.99\columnwidth]{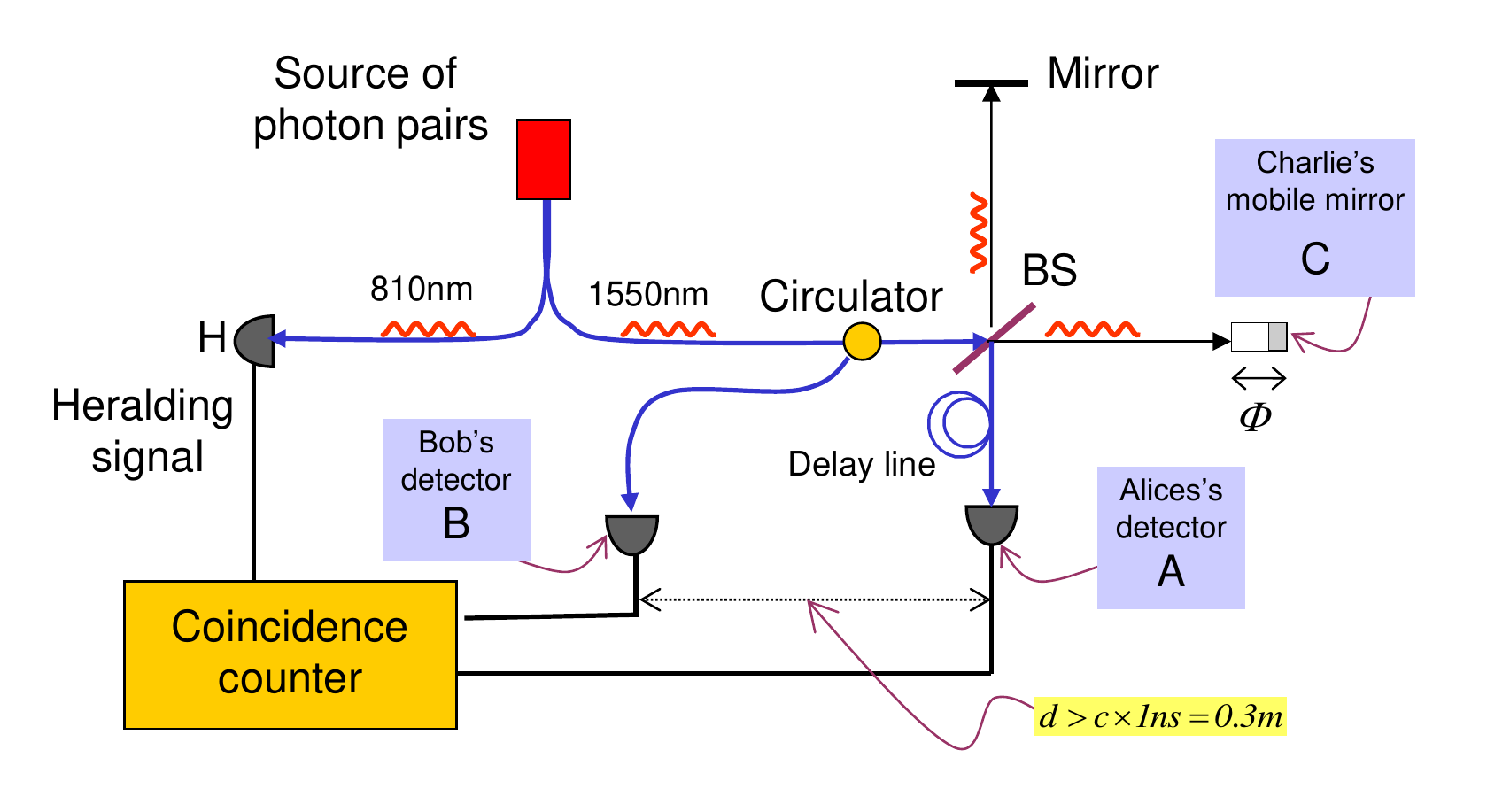}
\caption{\textbf{Single-photon space-like Michelson-Morley experiment:} The assumption of a time-ordered (non-covariant) collapse at the detectors A and B implies that one can define the velocity of light in an absolute way with relation to a preferred referential frame, and this implies a positive result, that is: the detection counting rates should change when the interferometer is rotated by $90\,^{\circ}$ ( see text).}
\label{f1}
\end{figure}

But what about Einstein's claim that there is ``a contradiction" between quantum nonlocality and relativity? In this paper we address this question and propose a new experiment that proves Einstein wrong: The standard postulate of the ``wavefunction collapse" can be interpreted as happening without any time-order and preferred frame, so that it not only does not contradict relativity, but rather implies it. To this aim, the setup of \cite{Guerreiro12} is completed with an interferometer as sketched in Figure \ref{f1}. This makes it possible to perform a new version of the Michelson-Morley experiment \cite{MM} that uses single-photons and two detectors (instead of only one). In this context the assumption of a ``time-ordered collapse'' means that one of the two detectors decides first to fire or not to fire, and determines how the other detector behaves thereafter. And this means that it is possible to define the velocity of light in an absolute way with relation to a preferred referential frame. Therefore the confirmation of relativity by a refreshed negative Michelson-Morley result is also a falsification of the time-ordered formulation of nonlocality at detection, and we argue that it may even be considered to refute Bohm's ``preferred frame" assumption as well.

\ \\
\textbf{Time-ordered collapse}.\textemdash Consider the experiment depicted in Figure \ref{f1}: A source produces pairs of photons and one of them is used for heralding, i.e. to signaling the presence of a photon in the interferometer and opening the counting gate, as indicated in \cite{Guerreiro12}. The other photon enters the interferometer through the beam-splitter BS and, after reflection in the mirrors, leaves through BS again and gets detected. Each interferometer's arm is supposed to have equal length $L$.

Such interference experiments can be considered the entry into the quantum world. With sufficiently weak intensity of light, only one of the two detectors clicks: either A or B (\emph{photoelectric effect}). Nevertheless, for calculating the counting rates of each detector one must take into account information about the two paths leading from the laser source to the detector (\emph{interference effect}). If $a \;\in\{+1,-1\}$ labels the detection value, according to whether detector A or detector B clicks, the probability of getting $a$ is given by:

\begin{footnotesize}
\begin{eqnarray}
P(a)=\frac{1}{2}(1+ a \cos \mathit\Phi)
\label{Pa}
\end{eqnarray}
\end{footnotesize}

\noindent where $\mathit\Phi=\omega\tau$ is the phase parameter, $\omega$ the angular frequency, and $\tau$ the optical path difference i.e., the difference between the times light take to travel each path of the interferometer. If one assumes the same light speed in the two paths of the interferometer, then $\tau=\frac{l-s}{c}$. The phase $\mathit\Phi$ can be changed by means of a mobil mirror C.

As said, according to standard quantum mechanics which detector clicks (the outcome) is decided by a choice on the part of nature when the information about the two paths reaches the detectors. And this implies a coordinated behavior on the part of A and B, no matter how far away from each other these detectors are.

On the other hand the experiment in Figure \ref{f1} can be used to a single-photon version of the Michelson-Morley experiment, which in 1887 provided the observational basis for relativity.

Suppose now that one describes the coordination between the two detectors A and B as a time-ordered influence fulfilling the two following conditions:

a) it happens with infinite speed,

b) the decision at one of the detectors happens first and determines (``is the cause of") the decision at the second detector.

The condition a) implies that the time-order invoked in b) can only arise from the assumption that the light needs more time to travel from the source to detector A than to detector B, or viceversa. And this assumption requires that the velocity of light can be defined in an absolute way, with relation to some ``preferred frame''. Accordingly one can expect that the reproduction of the Michelson-Morley experiment with the setup of Figure \ref{f1} gives a positive result (i.e. shows a shift of the counting rates of A and B). If by contrast the experiment upholds frame independence for the speed of the light, this result implies frame independence for the collapse as well, or in other words, it rules out the time-ordered description of the quantum collapse and confirms the timeless one.

\ \\
\textbf{The single-photon space-like Michelson-Morley experiment}.\textemdash Consider the experiment of Figure \ref{f1} with one of the arms oriented in the direction of the Earth's motion. According to the ``preferred frame'' assumption, if the Earth moves with velocity $v$ relative to the ``preferred frame'',  the travel time of the light through each interferometer's arm exhibits a difference given by:

\begin{footnotesize}
\begin{eqnarray}
\tau_1= L \frac{v^2}{c^3}
\label{tau1}
\end{eqnarray}
\end{footnotesize}

By rotating the interferometer 90$^\circ$ one interchanges the orientation of the two paths relative to the ``preferred frame'' and gets the following travel time difference for the light:

\begin{footnotesize}
\begin{eqnarray}
\tau_2= - L \frac{v^2}{c^3}
\label{tau2}
\end{eqnarray}
\end{footnotesize}

Taking account of (\ref{tau1}) and (\ref{tau2}) the phase shift after rotation of the interferometer is given by:

\begin{footnotesize}
\begin{eqnarray}
\Delta\mathit\Phi =2 \pi \nu (\tau_1 - \tau_2) =4\pi \frac{c}{\lambda} L \frac{v^2}{c^3}=4\pi\frac{L}{\lambda}\frac{v^2}{c^2}
\label{MM2}
\end{eqnarray}
\end{footnotesize}

\noindent where $\nu$ means the frequency and $\lambda$ the wavelength of the photons.

Like in the original Michelson-Morley experiment \cite{MM} we take $v=30 km/s$ as the orbital velocity of the Earth around the Sun \cite{mm}, and assume that at some moment of the day (because of the Earth rotation) one of the arms will be oriented near to the direction of the ``preferred frame''. According to \cite{Guerreiro12} we assume photons of wavelength about $\lambda=1550nm$.

Taking account of the results in \cite{Guerreiro12} for the counting rates in each of the detectors A and B and the noise, we try to get a well testable prediction by setting parameters such that the counting rate changes from $P_A=0.50$ to $P_A=0.25$ in Alice's detector, and from $P_B=0.50$ to $P_B=0.75$ in Bob's one when the interferometer is rotated by 90$^\circ$. According to (\ref{Pa}) this counting rate shift corresponds to a phase shift of $\Delta\mathit\Phi=\frac{\pi}{6}$, and from (\ref{MM2}) one gets the following length $L$:

\begin{footnotesize}
\begin{eqnarray}
4\pi\frac{L}{\lambda}\frac{v^2}{c^2}=\frac{\pi}{6}\;\; \Rightarrow \;\;\; L = 6.25 m
\label{MM3}
\end{eqnarray}
\end{footnotesize}

Hence, if one uses an interferometer fulfilling (\ref{MM3}), the ``preferred frame'' theory predicts that at some moment of the day the counting rates of the detectors A and B fulfill:

\begin{footnotesize}
\begin{eqnarray}
&&P_B-P_A=0,\;\;\; \texttt{before rotation of 90$^\circ$} \nonumber\\
&&P_B-P_A=0.50, \;\;\;\texttt{after rotation of 90$^\circ$}
\label{MM4}
\end{eqnarray}
\end{footnotesize}

Equation (\ref{MM4}) provides a clear trial of the ``preferred frame'' assumption.

We denote $R_{HA}$ the total number of coincident counts at detector H and detector A during the time of measurement, and $R_{H(A)}$ the total number of counts at detector H alone during the same measurement; $R_{HB}$ and $R_{H(B)}$ denote similar quantities for the measurement with H and B. All these quantities can directly be obtained by measurement and are related to the probabilities in (\ref{MM4}) by the equations:

\begin{footnotesize}
\begin{eqnarray}
P_A = \frac{R_{HA}}{R_{H(A)}},\;\;\;P_B = \frac{R_{HB}}{R_{H(B)}}
\label{PR1}
\end{eqnarray}
\end{footnotesize}

\noindent therefore the prediction of the ``preferred frame'' assumption (\ref{MM4}) can be tested by experiment.

Simultaneity of detection at A and B can be guaranteed to within 1ns by the matched length of fiber both before and inside the detectors. Thus if the detectors A and B are separated by $d> 0.3m$, the corresponding decisions at A and B are space-like separated \cite{Guerreiro12}. In accord with the results in \cite{Guerreiro12} we assume that there are no triple coincident counts at the detectors H, A and B, and therefore A and B behavior requires nonlocal coordination. Then falsification of the ``preferred frame'' assumption by the experiment of Figure \ref{f1} rules out the time-ordered collapse, and upholds at the same time the relativity. And on its turn relativity implies that the timeless quantum collapse is nonlocal and comes from outside the space-time.

\ \\
\textbf{Discussion}.\textemdash This experiment (Figure \ref{f1}) has relevant implications regarding the interpretation of quantum mechanics:

\ \\
\emph{Non-covariant and covariant collapse}.\textemdash Suppose Charlie changes the settings of the mobile mirror C in the setup of Figure \ref{f1}, and consequently the firing rates of the detectors A and B. Thereby he produces a correlation between the mirror's settings and the behavior of the detectors, and can send a message to Alice and Bob. Such a procedure is a paramount example of relativistic causality that can be used for communication through signaling (e.g.: phoning). When using the term ``Lorentz invariance'' or ``covariance'' one should distinguish two different meanings:

1) The observed correlations and the corresponding relation cause-effect are independent of the observer's inertial frame.

2) The time-order between two causally linked events is the same in any inertial frame, that is for any observer.

The meaning 2) reveals the widespread cognitive tendency to tie causality to time, which  in my opinion is at the origin of the prejudice that relativity and quantum mechanics are ``incompatible''. Like John Bell I wholeheartedly claim that ``correlations cry out for explanation'', and consider correlation as the very sign of causality. Nonetheless I disconnect causality from time-order.

So (in line with \cite{bde}, p. 331) I consider that both, the \emph{correlation} between the settings of the mobile mirror C and the counting rates of the detectors A and B, and the \emph{correlation} between the two detectors A and B, reveal a \emph{causal relation}. While the former can be linked to a time-order and used by Charlie for phoning to Alice or Bob, the second one comes from outside the space-time and cannot be controlled by Alice and Bob for phoning. \emph{In both cases the correlation is independent of the observer's inertial frame.}

Accordingly I suggest to use the term ``covariance'' only with the meaning 1) above, and consider ``covariant'' both, the \emph{relativistic local} correlation between the mirror and each detector, and the \emph{quantum nonlocal} correlation between the two detectors A and B. By contrast the time-ordered picture of the collapse (tying causality to time) is non-covariant. In this sense the experiment of Figure \ref{f1} shows that both relativity and quantum mechanics are ``covariant" or share ``Lorentz invariance'', and test this feature against the proposed non-covariant nonlocal collapse.

To complete the analysis it is interesting to discuss the possibility of defining a time-ordered collapse using different frames, instead of a unique absolute preferred frame. This is what achieves the Multisimultaneity model, which combines time-ordered causality and the relativity of simultaneity by using the frames of the measuring devices \cite{asvs97}: Suppose that in in the experiment of Figure \ref{f1} one sets the two detectors moving away from each other so that the decision at detector A in the inertial frame of A happens before the decision at B, and the decision at B, in the inertial frame of B, happens before the decision at A (before-before timing). Since each detector's decision cannot be determined by the decision at the other, the model predicts the disappearance of the nonlocal correlation leading to the violation of the conservation of energy in the single quantum events. This prediction can be considered ruled out by the experiment in \cite{Guerreiro12}.

\begin{figure}
\includegraphics[trim = 2mm 160mm 0mm 36mm, clip, width=0.90\columnwidth]{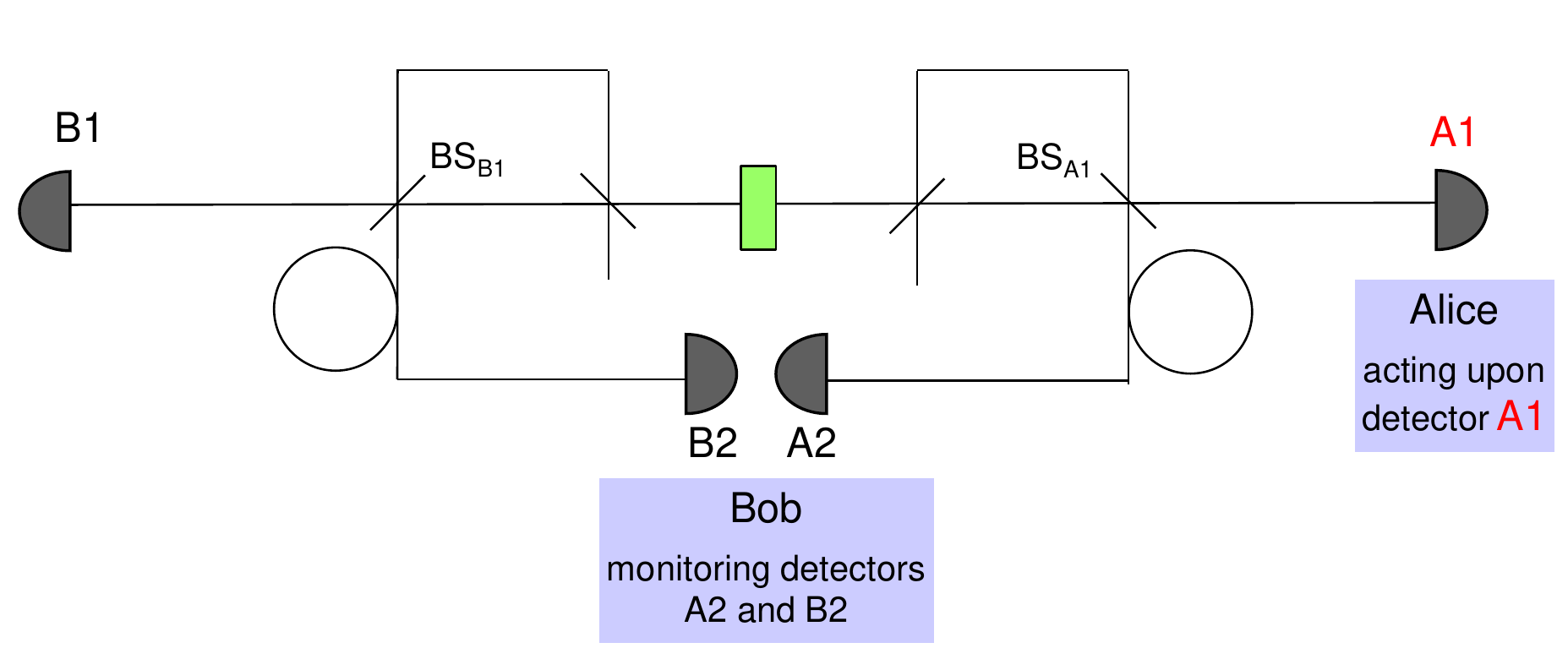}
\caption{\textbf{2-particle entanglement experiment with moving detector:} The attempt to order the quantum collapse using the referential frames of the detectors (see text) conflicts with relativity: A1 in red means that Alice can break the nonlocal coordination between detector A1 and detector B2 by changing the state of movement of A1 to set up before-before timing. Then the rate of the joint outcomes (A2, B2) depends on the settings of detector A1, and Alice (operating upon A1) can message to Bob (watching at B2 and A2) faster-than-light.}
\label{f2}
\end{figure}

Additionally, by means of the thought experiment sketched in Figure \ref{f2} (which reproduces an argument by Andr\'{e} Stefanov included in \cite{ZBGT01}) one shows that any arrangement of the detectors thwarting the nonlocal coordination between Alice's detectors and Bob's ones would allow Alice phoning to Bob faster-than-light and therefore conflicts with relativity.

\ \\
\emph{Non-covariant ``pilot wave"}.\textemdash To avoid the standard quantum nonlocality at detection in single-particle experiments Einstein invoked in 1927 Louis de Broglie's explanation by means of ``particle and pilot wave" \cite{bv}, which implies that the outcome is determined at the beam-splitter.

Nonetheless, de Broglie's local picture cannot be extended to entanglement experiments with two or more particles. Einstein himself is at the origin of this insight with the celebrated EPR paper in 1935: The local hidden variables of de Broglie (particle and pilot wave) do not account for the nonlocal EPR correlations quantum mechanics predicts \cite{jb}. To overcome this problem David Bohm completed the de Broglie's local model with a ``nonlocal quantum potential''. The theory was quite inspiring for John Bell since it highlights that local hidden variable models do not suffice to explain quantum mechanics. Bell showed that such models fulfill the well known locality criteria of Bell inequalities, and are refuted by the experimental violation of these in the conventional Bell experiments \cite{jb}.

Moreover, de Broglie's description cannot be considered ``realistic": The ``pilot wave'' is not a wave propagating in some material ether within the ordinary 3-space, but rather a mathematical entity defined in the so called ``3N-space or configuration space" (\cite{jb} p. 128). In other words, the ``wave'' guides the particle from outside the ordinary space-time.

This means that de Broglie-Bohm's model neither restores realism nor locality at the end, and Bohm's completion looks somewhat like an \emph{ad hoc} solution merging quite heterogeneous ingredients. Indeed, a main oddity of the model is that the principle of nonlocality appears only in experiments with two or more particles but does not rule at all the single-particle experiments, whereas one would expect that so a fundamental feature like nonlocality pervades the whole quantum mechanics, and governs the single-particle phenomena too. If the motivation for keeping to Bohmian mechanics is to maintain a local realistic description to some extent, why not do this in a more consistent way and adopt a thorough local realistic view that holds also for many-particle experiments? This is definitely possible by entering the ``Church of the Large Hilbert Space'' and adhering to the version of ``many worlds'' called ``parallel lives''  \cite{brr}.

Another frequent defence of the model claims that it saves the idea of temporal causality: Bohm's ``nonlocal potential'', although traveling at infinite velocity, is supposed to produce time-ordered outcomes with relation to some ``preferred frame'' so that one outcome event is the cause, and the other the effect \cite{db,jb}. However the model does not define the very notion of ``preferred frame'' (its essential characteristic) in a clear and unambiguous way, and this omission is highly eloquent: If one assumes infinite-speed influences to explain the quantum correlations in the context of experiments with entangled photon pairs, how can one define a time-order of the events? The only possibility to do this it is to assume that the time-order comes from the arrivals of the photons at the corresponding beam-splitters: So for instance in the experiment sketched in Figure \ref{f2}, one could suppose that Alice's photon arrives at BS$_{A1}$ before Bob's photon arrives at BS$_{B1}$, and this way Alice's outcome is the cause and precedes Bob's one, which is the effect. But this assumption implies again that it is possible to define the speed of light in an absolute way. In other words, the ``preferred frame" Bohm's theory postulates cannot be other than that the experiment of Figure \ref{f1} aims to test, and Bohm's ``nonlocal potential'' reduces to the time-ordered quantum collapse.

Also here one could invoke Multisimultaneity and attempt to combine Bohm's idea of nonlocal time-ordered potential with the relativity of simultaneity by using the frames of the beam-splitters. Consider again the experiment of Figure \ref{f2} and suppose the two beam-splitters BS$_{A1}$ and BS$_{B1}$ are moving away from each other. Suppose that in the inertial frame of BS$_{A1}$ the decision at BS$_{A1}$ happens before the decision at BS$_{B1}$, and in the inertial frame of BS$_{B1}$,the decision at BS$_{B1}$  happens before the decision at BS$_{A1}$ (before-before timing). Since each beam-splitter's decision cannot be determined by the decision at the other, the model predicts that the nonlocal correlation should disappear under the ``before-before" configuration. This prediction has been ruled out by the experiment described in \cite{SZGS02}. More recently, Multisimultaneity has been proved to imply communication faster than light and conflict with relativity for 3- and 4-particle experiments \cite{Scarani13}. Note also that Multisimultaneity is non-covariant and cannot be formulated in a covariant way \cite{ng}.

In summary, when one tries to give a sharp formulation of the ``pilot wave" model it seems that one is led either to ``many worlds'' or to the non-covariant collapse presented in this paper, which intrinsically conflicts with relativity. In this sense the experimental refutation of the latter means at the same time a refutation of Bohm's ``nonlocal potential''.

\ \\
\textbf{Conclusion}.\textemdash The facts that special relativity arrived before quantum mechanics, and the Michelson-Morley experiment has never been done with single-photons and two detectors, have led to overlook that this experiment is basic not only for relativity but also for quantum mechanics. The experiment presented in in this paper (Figure \ref{f1}) closes this loophole bringing to light that the quantum collapse excludes any preferred frame, and in this sense implies ``Lorentz-invariance'' and relativity. By contrast, any sharp formulation of the quantum collapse in terms of time-ordered (non-covariant) infinite-speed influences appears to lead to (testable and falsifiable) predictions conflicting with relativity.

On the other hand by confirming relativity the experiment shows that the correlated behavior exhibited by A and B cannot be explained by influences propagating with velocity $v\leq c$ : The quantum collapse implies relativity, and the relativistic structure of the space-time implies that the quantum collapse is nonlocal (i.e., comes from outside the space-time) and cannot be used by the experimenter for communication faster than light.

Thus the proposed experiment clearly highlights that relativity and quantum physics are two inseparable aspects of one and the same description of the physical reality. These two theories neither are incompatible with each other nor have a ``frail peaceful coexistence'', but rather imply each other: we can't have one without the other.

\ \\
\emph{Acknowledgments}: I am thankful to Bernard d'Espagnat and Andr\'{e} Stefanov for extended comments and work in progress, and acknowledge also discussions with Nicolas Gisin, Thiago Guerreiro, Marc Lachi\`{e}ze-Rey, Carlo Rovelli, Bruno Sanguinetti, and Hugo Zbinden.


\begin{references}

\bibitem{bv} Guido Bacciagaluppi, Antony Valentini, Quantum Theory at the Crossroads: Reconsidering the 1927 Solvay Conference. Part III: The proceedings of the 1927 Solvay conference. Cambridge University Press, 2009. arXiv:quant-ph/0609184v2 (2009)


\bibitem{Guerreiro12} T. Guerreiro, B. Sanguinetti, H. Zbinden, N. Gisin, A. Suarez, Single-photon space-like antibunching, \emph{Phys. Lett. A} \textbf{376}, 2174–2177 (2012); A. Suarez, Empty waves, many worlds, parallel lives, and nonlocal decision at detection, arXiv:1204.1732 (2012).

\bibitem{jb} J. S. Bell, \emph{Speakable and Unspeakable in Quantum Mechanics}, Cambridge Univ. Press: Cambridge, 1987, 2004.

\bibitem{db} D. Bohm, A suggested interpretation of the quantum theory in terms of ``hidden'' variables. I and II. \emph{Phys. Rev.} \textbf{85} 166-193 (1952).

\bibitem{asvs97} A. Suarez, V. Scarani, Does entanglement depend on the timing of the impacts at the beam-splitters? \emph{Phys. Lett. A} \textbf{232}, 9-14 (1997); A. Suarez, Relativistic nonlocality in an experiment with 2 non-before impacts, \emph{Phys. Lett. A} \textbf{236}, 383-390 (1997), Quantum mechanics versus multisimultaneity in experiments with acousto-optic choice devices, {\em Phys. Lett. A}, {\bf 269}, 293 (2000).

\bibitem{ZBGT01} H. Zbinden, J. Brendel, N. Gisin, W. Tittel, Experimental test of nonlocal quantum correlation in relativistic configurations, \emph{Phys. Rev. A} \textbf{63}, 022111 (2001).

\bibitem{SZGS02} A. Stefanov, H. Zbinden, N. Gisin, A. Suarez, Quantum Correlations with Spacelike Separated Beam Splitters in Motion: Experimental Test of Multisimultaneity, \emph{Phys. Rev. Lett.} \textbf{88}, 120204 (2002); Quantum entanglement with acousto-optic modulators: Two-photon beats and Bell experiments with moving beam splitters, \emph{Phys. Rev. A} \textbf{67}, 042115 (2003).


\bibitem{Scarani13} V. Scarani, J.-D. Bancal, A. Suarez, and N. Gisin, Strong constraints on models that explain the violation of Bell inequalities with hidden superluminal influences, arxiv:1304.0532 (2013); A. Suarez, Any nonlocal model assuming "local parts" conflicts with relativity, arXiv:1304.4801

\bibitem{MM} A. A. Michelson and E. W. Morley,  On the Relative Motion of the Earth and the Luminiferous Ether, \emph{American Journal of Science} \textbf{34}, 333–345 (1887).

\bibitem{mm} R. S. Shankland, S. W. McCuskey, F. C. Leone, and G. Kuerti  New Analysis of the Interferometer Observations of Dayton C. Miller, \emph{Rev. Mod. Phys.} \textbf{27}, 167–178 (1955).



\bibitem{bde}Bernard d'Espagnat, \emph{On Physics and Philosophy}, Princeton University Press: Princeton, N.J. (2006).

\bibitem{brr} G. Brassard and P. Raymond-Robichaud, Can Free Will Emerge from Determinism in Quantum Theory? In: A. Suarez and P. Adams (Eds.) Is Science compatible with free will? Exploring free will and consciousness in light of quantum physics and neuroscience. Springer, New York, 2013, Chapter 4. arXiv:1204.2128v1.

\bibitem{ng} N. Gisin, On the Impossibility of Covariant Nonlocal ``hidden" Variables in Quantum Physics. Phys. Rev. A, \textbf{83}, 020102-1/2, (2011). arXiv:1002.1390 (2010).


\end{references}
\end{document}